# Enhancement of the superconducting critical temperature of $Sr_2CuO_{3+\delta}$ up to 95K by ordering dopant atoms

Q.Q.Liu, H.Yang, X.M.Qin, Y.Yu, L.X.Yang, F.Y.Li, R.C.Yu, C.Q.Jin*

*Institute of Physics, Chinese Academy of Sciences, P. O. Box 603, Beijing 100080, P. R. China*

S. Uchida*

*Department of Physics, University of Tokyo, 7-3-1Hongo, Bunkyo-ku, Tokyo 113-0033, Japan*

## Abstract

We address the question of whether the superconducting transition temperature ($T_c$) of high-$Tc$ cuprates is enhanced when randomly distributed dopant atoms form an ordered array in the charge reservoir layers. This study is possible for the $Sr_2CuO_{3+\delta}$ superconductor with $K_2NiF_4$-type structure in which oxygen atoms only partially occupy the apical sites next to the $CuO_2$ planes and act as hole-dopants. We show that remarkable $T_c$ enhancement up to 95K in this mono CuO2 layered HTS is associated with the apical oxygen ordering, not to the hole concentration change. The result points a route toward further enhancement of $T_c$ in cuprate superconductors.

# Introduction

One of the central concerns of high-temperature cuprate superconductor (HTS) is how to raise the superconducting transition temperature $T_c$. The doping level and the number of $CuO_2$ planes ($n$) in a unit cell have been considered to be most feasible parameters controlling $T_c$. For each member of cuprates, $T_c$ becomes maximum $T_c^{max}$ at optimal doping concentration, which is usually in the range between 0.15 and 0.20/Cu site[1], and $T_c$ is highest when $n=3$ for multilayer cuprates (or homologous series)[2]. HTS is an alternating array of charge reservoir blocks and the $CuO_2$ conducting planes. The charge reservoir blocks supply holes or electrons into the $CuO_2$ planes, and in most cases are disordered since dopant atoms reside in the blocks and are randomly distributed.

Tuning the oxygen content in the charge reservoir blocks is one of the most fundamental chemical doping mechanisms of HTS. As a typical example, by breathing in or out oxygen and forming the Cu-O chain layer, $YBa_2Cu_3O_{6+\delta}$ can be either a 90K superconductor or a non-superconducting insulator. In this case oxygen doping and its ordering is in the 2nd-nearest neighbor charge reservoir blocks, and it has been believed that disorder in these blocks has minimal effect on the electronic state in the $CuO_2$ planes as the relative long separation. On the other hand, the oxygen atoms at the apical sites of a $CuO_5$ pyramid or a $CuO_6$ octahedron called "apical oxygen" form the nearest neighbor charge reservoir block[2-4]. Disorder in this block is expected to have an appreciable effect on the adjacent $CuO_2$ plane since they have direct chemical bonding in terms of electron exchange interaction [5]. *However, for almost all HTSs, apical oxygen sites are fully occupied*[6]*, leading to the apical oxygen amount is unchangeable, not less to modulate the order state.* Study of a compound with partially occupied apical oxygen sites is thus of fundamental physical interest for a deeper understanding of the yet unresolved doping/order effect on high-$T_c$ superconductivity.

The $Sr_2CuO_{3+\delta}$ superconductor synthesized under high pressure is quite a unique cuprate which crystallizes into an oxygen-deficient $La_2CuO_4$ (*i.e.* $K_2NiF_4$) structure with partially occupied apical sites (Fig. 1), and the apical oxygen acts as hole dopant[6-10]. We have investigated the effect of low-temperature annealing on superconductivity in single phase $Sr_2CuO_{3+\delta}$, and found that a remarkable enhancement of $T_c$ in this superconductor is associated with the ordering of the apical oxygen atoms, not with a change of doped hole density.

At ambient pressure, the stoichiometric $Sr_2CuO_3$ forms an orthorhombic structure with Cu-O chains along the *a*-axis [14,15]. Introducing extra O and applying high pressures leads to the formation of a $K_2NiF_4$-type tetragonal structure containing a $CuO_2$ plane as shown in Fig. 1. Although high-pressure synthesis is a very powerful tool to search for novel HTS materials[6-12], preparation of a single-phase sample is general difficult, and sometimes it is hard to identify the superconducting phase in the sample [16,17]. Using $KClO_4$ as an oxidizer, Hiroi *et al.* [8] succeeded in fabricating tetragonal $Sr_2CuO_{3+\delta}$ showing superconductivity at $T_c = 70$ K. They suggested that the main phase in this material is a highly apical oxygen-deficient $K_2NiF_4$-type tetragonal structure. After this discovery $Sr_2CuO_{3+\delta}$ superconductors were synthesized by several groups using the high-pressure technique as well as thin-film growth [18], and $T_c$ was enhanced above 90 K by post annealing [9,10].

In the previous studies, two types of modulated structures were reported. Hiroi *et al.* [8] and Laffez *et al.* [9] found $4\sqrt{2}\ a_p \times 4\sqrt{2}\ a_p \times c_p$ and $5\sqrt{2}/2a_p \times 5\sqrt{2}/2a_p \times c_p$ modulated structures in their as-prepared samples, and suggested that the superconductivity might be occurring in the latter tetragonal phase. This was also supported by Wang *et al.* [19] and Zhang *et al.* [20]. However, these superconducting samples were multi-phase mixtures and superconducting volume fraction was very small, so it was difficult to determine the detailed structure and to identify the superconducting phase definitely.

## Experimental

To make the phase as pure as possible, we synthesized the $Sr_2CuO_{3+\delta}$ sample under pressures using $SrO_2$ as an oxidizer. The high-pressure synthesis (6GPa and 1100°C for 1hour) was performed using a cubic-anvil-type apparatus. $Sr_2CuO_3$, $SrO_2$, and CuO were mixed to yield the nominal composition $Sr_2CuO_{3+\delta}$ at various molar ratios in a dry box. The role of $SrO_2$ peroxide is to create an oxygen atmosphere during the high-pressure synthesis as we previously used in the related Sr-Ca-Cu-O-Cl system[11]. The oxygen pressure was controlled by the amount of $SrO_2$ in the starting materials. The main reason for using $SrO_2$ oxidizer instead of $KClO_4$ is to avoid the formation of the superconducting phase containing Cl as found in Ref. 17 or other unwanted pollution from the third element which will make the identification of real superconducting phase easier. The single phase $Sr_2CuO_{3+\delta}$ in terms of powder x-ray diffraction provides us the solid ground to make clear the origin of the mysterious high Tc in this mono layered $La_2CuO_4$ type superconductor. The crystal structure was analyzed by means of powder x-ray diffraction (XRD) using Cu K $\alpha$ radiation. The DC magnetic susceptibility was measured with a SQUID magnetometer in an external magnetic field of 20 Oe. A Tecnai F20 electron microscope with a field emission gun was used for electron diffraction (ED), HRTEM and EELS experiments at an acceleration voltage of 200 keV.

## Results and discussions

A series of $Sr_2CuO_{3+\delta}$ samples were prepared under high-pressures by changing the initial amount of $SrO_2$. The superconducting phase was found in the range $0.1 < \delta < 0.6$ and a maximum $T_c \approx 75$ K occurs at $\delta = 0.4$. The X-ray diffraction pattern of $Sr_2CuO_{3+\delta}$ ($\delta = 0.4$) shows tetragonal structure with space group $I4/mmm$, indicating an apparently single-phase pattern with lattice parameters $a = 3.795(3)$ Å, $c = 12.507(1)$ Å. In order to further "optimize" $T_c$, the sample was then annealed in the temperature range of between 150 and 350° C for 12h under 1 atm $N_2$ atmosphere in a tube furnace. The XRD pattern shows that the sample maintains the tetragonal single phase up to 300°C, and then becomes a mixture of the tetragonal and the orthorhombic chain phases when heat-treated at 350°C. Figure 2 presents the x ray diffraction patterns of $Sr_2CuO_{3+\delta}$ superconductor for both as-prepared and post annealed at 300°C.

Figure 3 shows the magnetic susceptibility measured for as-prepared and annealed sample in the Meissner (field-cooling) mode which reflects the superconducting volume fraction. It is found that with raising the annealing temperature to 200-250°C, $T_c$ increases to 95 K which is one of highest among the known single-layer cuprate superconductors, much higher than that of 39 K Tc in the isostruture compound $Sr_2CuO_{2+\delta}Cl_{2-x}$ superconductor[21]. Although similar $T_c$ values were reported previously on nominally the same compound[10], the present work confirms bulk superconductivity in single-phase samples[22]. We highlight that the post annealing temperature (150-250 °C) are too low to cause a change of oxygen content (which usually happens at temperatures above 300°C). At such low temperatures metal ions cannot migrate, either, so the effect of post-annealing is most likely to rearrange the apical oxygen atoms and make them to order. To confirm this, we performed a transmission-electron-microscope (TEM) study.

Electron diffraction (ED) and high-resolution TEM (HRTEM) investigations show that almost all grains in the superconducting samples exhibit modulated structures in the basal *a-b* plane. In the as-prepared sample, two types of modulated structures were found. Figures 4(a) show the ED patterns of the two modulated phases taken along $[001]_p$ zone-axis. One is the previously reported face-centered orthorhombic modulated structure (space group *Fmmm*) of about $5\sqrt{2}\,a_p \times 5\sqrt{2}\,a_p \times c_p$ periodicity[9]. The other is a new base-centered monoclinic modulated structure (space group *C2/m*) with the unit-cell parameters $a = 5\sqrt{2}\,a_p$, $b = c_p$, $c = \sqrt{26}\sqrt{2}/2 a_p$ and $\beta = 101.3^o$. We investigated 50 grains to estimate the relative fraction of the two phases. Statistically about 80% of the grains have the *Fmmm* structure, and the rests are the *C2/m* phase.

After heat treatment at 150 $^o$C, 15% of the *C2/m* modulated phase was found to convert to another new modulated phase (space group *Cmmm*) with the unit-cell parameters $a = c_p$, $b = 5\sqrt{2}\,a_p$ and $c = 5\sqrt{2}\,a_p$, while the *Fmmm* modulated structure is unchanged keeping almost the same volume fraction. Figure 4(b) shows the ED pattern along $[001]_p$ of the new modulated phase (*Cmmm*). The phase transformation from *C2/m* to *Cmmm* can also be seen in the HRTEM images. The HRTEM image (Fig. 3(b)) shows coexistence of the two modulated phases. The domain A and B correspond to the *C2/m* and the *Cmmm* modulated phases, respectively.

When the annealing temperature was increased to 250 $^o$C, all the *C2/m* and *Cmmm* modulated phases were converted to a single modulated phase (space group *Pmmm*) with the unit-cell parameters $a \approx b = 4\sqrt{2}\,a_p$ and $c = c_p$ which was previously reported by

Hiroi *et al.*[8]. Figure 4(c) displays the ED pattern along $[001]_p$ of the *Pmmm* modulated phase. Again, the *Fmmm* modulated phase remains in the sample and continues to occupy 80% volume of the sample.

The results of our experiments strongly suggest that the observed superconductivity is related to those modulated phases. The as-prepared sample with $T_c$=75 K contains two types of modulated phases with space groups *Fmmm* and *C2/m*, so either *Fmmm* or *C2/m* modulated phase is superconducting. We see that the *Fmmm* modulated phase always exists as a major phase in both as-prepared and annealed superconducting samples. This means that, if the *Fmmm* modulated phase were superconducting, nearly the same $T_c$ (~75 K) could have been observed for the annealed samples. Therefore, the *Fmmm* modulated phase is not superconducting, and the *C2/m* modulated phase is a superconductor with $T_c$ = 75K. The relatively small superconducting fraction (15-20%) estimated from the susceptibility (Fig. 3) also supports this conclusion. These observations indicate that the superconductivity or $T_c$ of $Sr_2CuO_{3+\delta}$ evolves with change of modulated structure: Starting from $T_c$ = 75K of the *C2/m* modulation structure in an as-prepared sample, $T_c$ goes up to 89 K in the *Cmmm* modulation structure after the 150°C annealing, and finally $T_c$ = 95 K is attained in the *Pmmm* modulation structure of the 250°C-post annealed sample. The *Pmmm* modulated phase is, in this respect, the *optimally* modulated phase. This resolves the mystery/controversy associated with this material [6-10 16-18].

As regards the relationship between oxygen vacancies and the observed modulation structures, the thermo-gravimetry (TG) analysis on the $Sr2CuO_{3+\delta}$ superconductor found little weight change below 300°C as shown in Fig.5, indicating a negligible change of oxygen content and hence the doping level after annealing. Since the modulation is along the *a-b* plane directions, its location is most likely on the rock salt type SrO charge reservoir block, and the modulation would be induced by the apical oxygen rearrangement / ordering [23]. It is consequently inferable that the apical oxygen ordering has a substantial effect on the superconducting transition temperature. The oxygen ordering has been well established in YBa$_2$Cu$_3$O$_{6+\delta}$ even at room temperature [24] but it takes place at the chain layer, i.e. the second nearest neighbor in charge reservoir block. Here the ordering in Sr2CuO$_{3+\delta}$ superconductor is at the apical oxygen layer, which locates at 1$^{st}$ nearest layer.

Then the question that arises is why the apical oxygen ordering has such a significant effect on $T_c$. The LDA band calculation illustrates that the distance $d_a$ of apical oxygen with respect to the CuO$_2$ plane has a substantial effect on the 2nd-nearest neighbor hopping integral *t'* between nearest Cu-Cu atoms in the CuO$_2$ plane[5] and *t'* has a correlation with the maximum $T_c$ of each cuprate material. The LDA band calculation[5] and also the evaluation of the Madelung potential difference between apical and planar O sites [25] find a general trend that $T_c$ becomes higher as |*t'*| becomes larger or as the apical oxygen distance $d_a$ becomes longer. In this respect, the CuO$_2$ plane without apical oxygen atoms like that in T'-Nd$_2$CuO$_4$ would sustain highest $T_c$ if it could be doped with *holes*. Normally, it is hard to dope holes into such CuO$_2$ planes due to relatively electro-positive circumstances around the CuO$_2$ plane, but an exceptional situation is realized in the multilayered cuprates with the consecutive CuO$_2$ plane number *n* larger than 3 (inclusive). In the multilayer systems, the inner CuO$_2$ planes have no apical oxygen and thus fewer hole density compared with the two outer layers [26]. It might be that the outer layers supply a sufficient density of holes, while the inner layers provide a place for strong pairing correlation, both working cooperatively to enhance $T_c$. In analogy with this we suppose that a similar situation may be realized in $Sr_2CuO_{3+\delta}$ in which the $CuO_4$ plaquettes with apical oxygen and those without apical oxygen form some ordered structure within the same $CuO_2$ plane. Further, the ordering of apical oxygen atoms would minimize the disorder effect, and probably enhance $T_c$ in this monolayer cuprate.

The high-pressure synthesis and the subsequent quenching to ambient pressure would leave some residual strains in the lattice. The annealing at low temperatures is expected to reduce or relax these strains by the apical oxygen relocation. The release of strain was suggested by Attfield *et al.* to account for the disorder effects on $T_c$ in the cation-substituted $La_2CuO_4$ type superconductor [27].

Normally, the chemical doping introduces disorder into the charge reservoir blocks owing to random distribution of dopant atoms. It has been suggested that the disorder might be responsible for the observed electronic inhomogeneity on the nanometer scale in the $CuO_2$ plane[28-30]. It has also been demonstrated that in Bi2212 and Bi2201, intentionally introduced cation disorder in the nearest neighbor (SrO) block containing apical oxygen sites gives rise to an appreciable decrease in $T_c$ [28-30]. The present system is a positive example in which the ordering of dopant atoms enhances $T_c$. Application of this method to other cuprate will lead to further enhancement of $T_c$ of cuprate superconductors.

## Acknowledgement:

This work was partially supported by the NSF and MOST of China through the research projects. S. U. was supported by a Grant-in-Aid for Scientific Research in Priority Area from MEXT, Japan. The authors would like to thank Prof. J. S. Zhou, H. Yamauchi and B. Raveau for inspired discussions.

**Figure captions**

Fig.1 Schematic view of the crystal structure of $Sr_2CuO_{3+\delta}$ with $K_2NiF_4$ type tetragonal structure containing the $CuO_2$ plane, but the apical oxygen sites are partially occupied.

Fig. 2 X-ray powder diffraction pattern for the sample $Sr_2CuO_{3+\delta}$ with nominal $\delta =0.4$: (a) the high pressure as prepared sample, (b) the sample after post heat treatment at 300°C in N2 at ambient pressure.

Fig. 3 Temperature dependence of the DC magnetic susceptibility in the field-cooling mode for as-prepared $Sr_2CuO_{3+\delta}$ and those after annealed at various temperatures in the $N_2$ atmosphere.

Fig. 4. (a) ED patterns taken along the $[001]_p$ zone-axes of the *Fmmm* modulated phase (left) and the *C*2/*m* modulated phase (right). The latter is superconductor with $T_c = 75K$.

(b) (Left) ED pattern taken along the $[001]_p$ zone-axis of the *Cmmm* modulated phase observed in the sample of Tc 89 K. (Right):HRTEM image along the $[001]_p$ direction showing the coexistence of the *C*2/*m* (domain A) and the *Cmmm* modulated phase (domain B), which is superconducting with $T_c=89K$.

(c) ED pattern taken along the $[001]_p$ zone-axis of the *Pmmm* modulated phase in the sample (heat treated at 250 °C), showing superconducting transition at 95K.

Figure 5. The thermo-gravimetry (TG) analysis on the Sr2CuO$_{3+\delta}$ superconductor, showing no weight loss below 300°C.

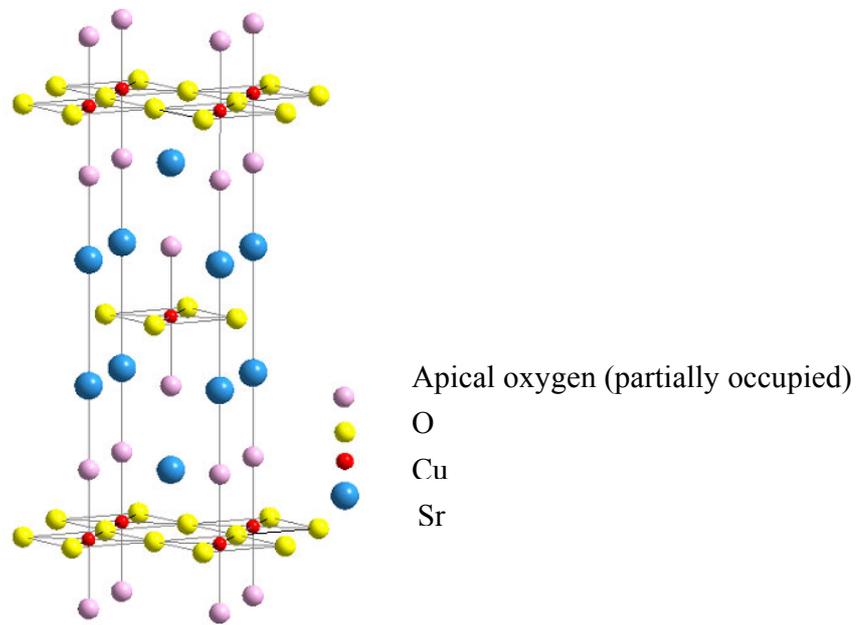

**Fig.1**

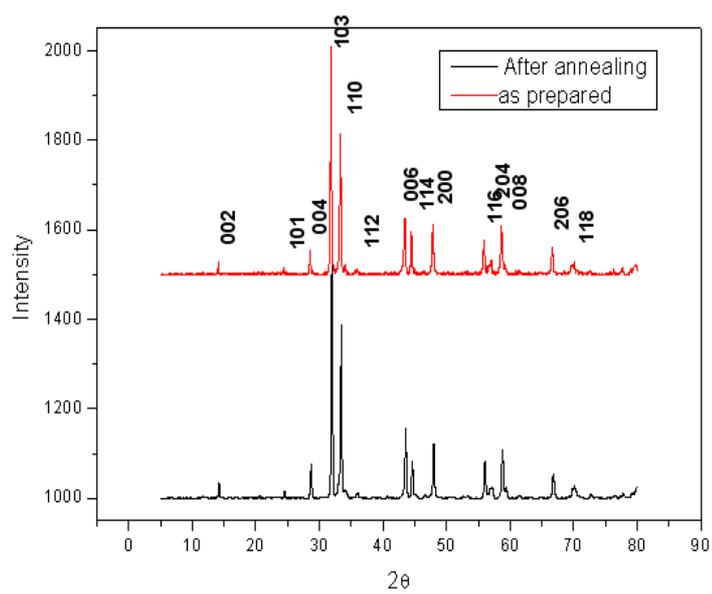

**Fig.2**

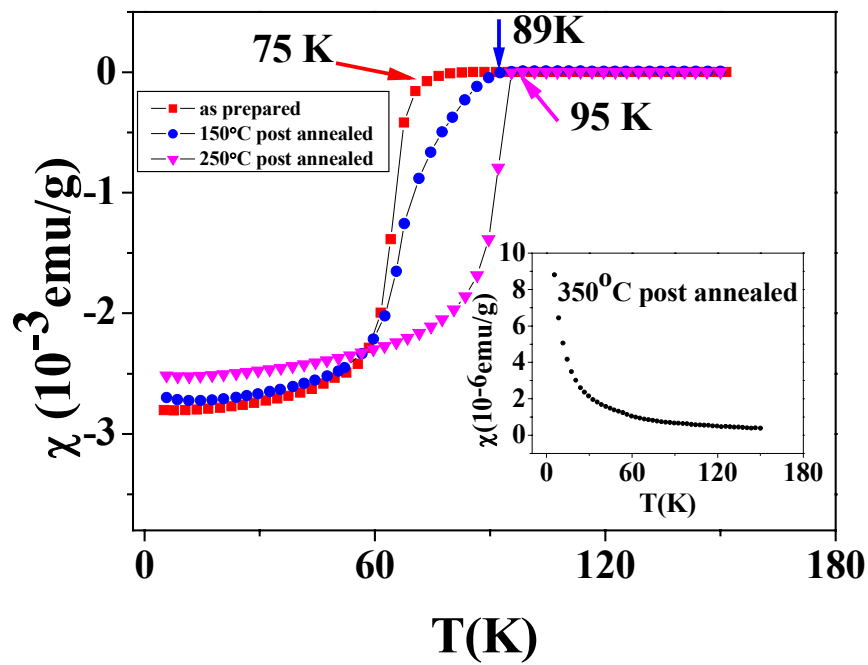

Fig.3

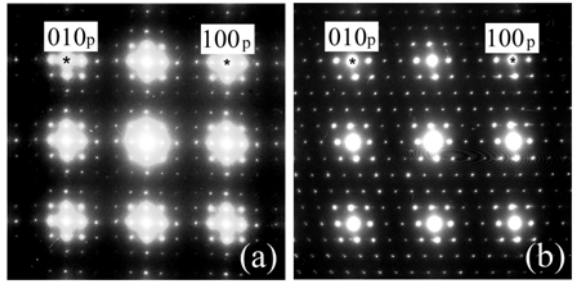

**Fig.4(a)**

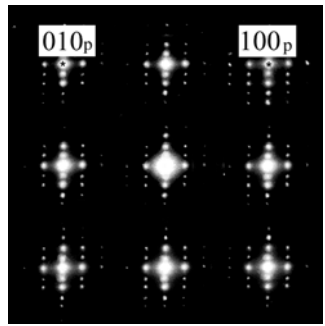

**Fig.4(b)**

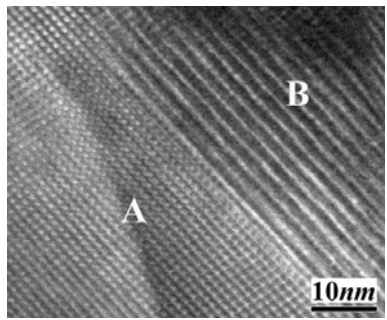

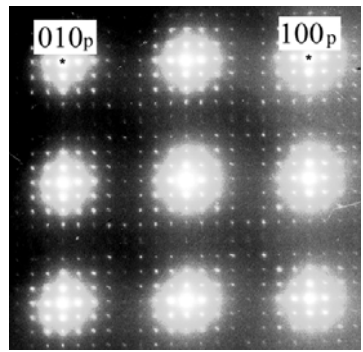

**Fig.4(c)**

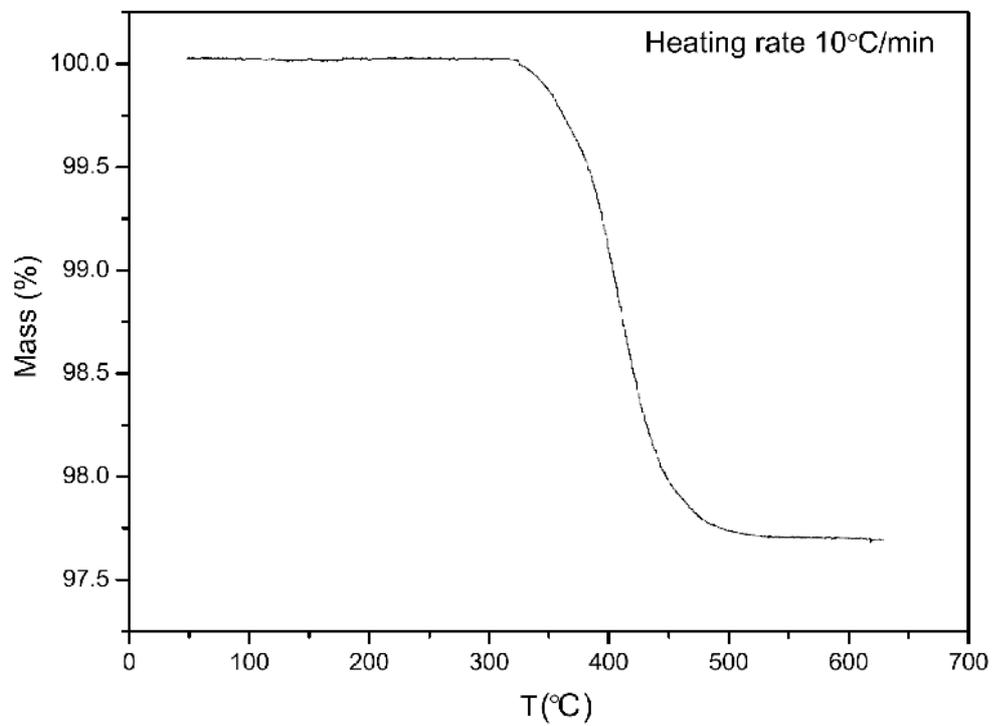

**Fig. 5**